\documentclass[oldversion]{aa}  

\usepackage{graphicx}
\usepackage{epsfig}
\usepackage{natbib}

\usepackage{txfonts}
\begin{document}
   \title{Strong [CII] emission at high redshift
 \thanks{Based on observations made with ESO telescope APEX,
 under program ID E-082.B-0692A-2008.}
}

   \author{R. Maiolino\inst{1}
          \and
          P. Caselli\inst{2}
		  \and
		  T. Nagao\inst{3}
		  \and
		  M. Walmsley\inst{4}
		  \and
		  C. De Breuck\inst{5}
		  \and
		  M. Meneghetti\inst{6,7}
          }

   \institute{
   INAF-Osservatorio Astronomico di Roma, via di Frascati 33, 00040
              Monte Porzio Catone, Italy
	\and
	School of Physics and Astronomy, University of Leeds, Leeds LS2 9JT, UK
	\and
	Research Center for Space and Cosmic Evolution, Ehime University,
	2-5 Bunkyo-cho, Matsuyama 790-8577, Japan
	\and
	Osservatorio Astrofisico di Arcetri,
	Largo E. Fermi 5, 50125 Firenze, Italy
	\and
	European Southern Observatory,
	Karl Schwarzschild Strasse 2, 85748 Garching, Germany
	\and
	INAF-Osservatorio Astronomico di Bologna, via Ranzani 1, 40127
	    Bologna, Italy
	\and
	INFN, Sezione di Bologna, viale Berti Pichat 6/2, I-40127 Bologna, Italy
             }

   \date{Received ; accepted }

 
  \abstract
   {We report the detection of the 
[C{\sc II}]157.74$\mu$m fine-structure line
in the lensed galaxy BRI 0952-0115 at z=4.43, using the
APEX telescope. This is the first detection of the [C{\sc II}] line 
in a source with  $\rm L_{FIR} < 10^{13} \, L_\odot$ at high
redshift. The line is very strong compared to previous [C{\sc II}]
detections at high-z
(a factor of 5--8 higher in flux), partly due to the lensing
amplification.
The $\rm L_{[CII]}/L_{FIR}$ ratio is $\rm 10^{-2.9}$, which
is higher than observed in local galaxies with similar
infrared luminosities. Together with previous observations of [C{\sc II}]
at high redshift, our result suggests that the [C{\sc II}] emission
in high redshift galaxies is enhanced relative to local galaxies of
the same infrared luminosity. This finding may result from
selection effects of the few current observations of [C{\sc II}] at high
redshift, and in particular the fact that non detections may have not
been published (although the few published upper limits are still consistent
with the [C{\sc II}] enhancement
scenario). If the trend is confirmed with larger samples,
it would indicate that high-z galaxies are
characterized by different physical conditions with respect to their
local counterparts. Regardless of the physical origin of the trend,
this effect would increase the potential of the [C{\sc II}]158$\mu$m
line to search and characterize high-z sources.
   }
   {}


   \maketitle
%

\section{Introduction}

The $^2P_{3/2} \rightarrow {^2P_{1/2}}$ fine-structure line of $\rm C^+$
at 157.74~$\rm \mu m$ is emitted predominantly by
gas exposed to ultraviolet radiation in
photo dissociation regions (PDRs) associated with star forming activity
(even in galaxies hosting AGNs).
This line is generally the brightest emission line in the
spectrum of galaxies, accounting for as much as $\sim$0.1-1\% of
their total luminosity \citep[][]{crawford85,stacey91,wright91}.
As a consequence, the [CII]158$\mu$m line is regarded as the most
promising tool to detect and identify high redshift galaxies with
forthcoming (sub-)mm facilities, such as ALMA
\citep[e.g.][]{maiolino08}.

In local galaxies, with far-infrared luminosities $\rm L_{FIR} < 10^{11}
\, L_\odot$, the [CII] luminosity is proportional to the far-IR luminosity,
and typically 
$\rm -3 \le log (L_{[CII]}/L_{FIR}) \le -2$ \citep[e.g.][]{stacey91}.
However, for sources with $\rm L_{FIR} > 10^{11}-10^{11.5} \, L_\odot$
this ratio drops by an order of magnitude
\citep[][]{malhotra01,luhman98,luhman03,negishi01}.
Various explanations have been proposed for the physical origin of 
this effect, more specifically:
1) a high ratio of ultraviolet flux to gas density, which results
in positively charged dust grains that in turn reduce the efficiency
of the gas heating by the photoelectric effect \citep[e.g.][]{kaufman99},
and which also increases the fraction of UV radiation absorbed by dust
\citep{abel09};
2) opacity effects which weaken the [C{\sc II}] emission
line in infrared luminous galaxies
\citep[][]{gerin00,luhman98}; 3) contribution to $\rm L_{FIR}$ from
dust associated with HII regions \citep[][]{luhman03}; 4) contribution to
$\rm L_{FIR}$ from an AGN \citep[see][]{malhotra01}.
Regardless of the physical origin of the effect, the sharp drop
of the $\rm L_{[CII]}/L_{FIR}$ ratio at high luminosities has cast
doubts about the usefulness of the [CII] to trace high-z galaxies.

Although locally the [CII]158$\mu$m line can only be observed from
space or from airborne observatories, at high redshift the line
is shifted into the submm-mm windows of atmospheric transmission,
and therefore can be observed with groundbased facilities.
The first detection of the [CII]158$\mu$m was obtained in
J114816.64+525150.3, one of the most distant quasars known at z=6.4,
for which the [CII] line is shifted at 1.2~mm \citep{maiolino05,walter09}.
The second most distant [CII] detection was obtained in
BR 1202--0725, a quasar at z=4.7 \citep{iono06}. Both these sources
have luminosities $\rm L_{FIR} > 10^{13} \, L_\odot$ (Hyper Luminous
Infrared Galaxies, HyLIRGs)\footnote{Note that the
``standard definition'' of LIRG, ULIRG and HyLIRG
is based on the total infrared luminosity ($\rm L_{IR} = L(8-1000)\mu m$),
and specifically the three classes are defined with
$\rm L_{IR}>10^{11},~>10^{12},~>10^{13}~L_{\odot}$, respectively
\citep{sanders96}.
Throughout the paper we instead use the far-IR luminosity,
defined as $\rm L_{FIR} = L(40-500)\mu m$, which is generally more accurately
constrained and reported for high-z sources and which can be
lower than $\rm L_{IR}$ by a factor ranging from 1.1 to 2 depending on
the source.\label{foot_firdef}},
and actually these are the first [CII]
detections in galaxies with such high luminosities. In these objects  
$\rm log (L_{[CII]}/L_{FIR}) \approx -3.5$, i.e. similar to the
value observed in local Ultraluminous Infrared Galaxies (ULIRGs) with
luminosities $\rm 10^{12} < L_{FIR} < 10^{13} \, L_\odot$. This result
is somewhat surprising, since the extrapolation of the rapidly
decreasing $\rm L_{[CII]}/L_{FIR}$ observed in local ULIRGs would
predict a much lower [CII] luminosity than observed in the two
extremely luminous objects at high-z detected so far. Possibly,
evolutionary effects may be involved.

To further investigate the latter issue one should search for [CII]
emission in high-z galaxies with infrared luminosity closer to the
local ULIRGs for which [CII] has been observed, i.e.
$\rm L_{FIR}\approx 10^{12}-10^{12.5} \, L_\odot$. The quasar BRI 0952-0115 
at z=4.4337 is an optimal source for this test. This is a lensed quasar
with a magnification factor in the range $\mu = 2.5-8$ (see Appendix),
whose mm and submm fluxes (fitted
with a modified black body)
indicate an intrinsic (de-magnified) far-IR luminosity
$\rm L_{FIR}\sim 4\times 10^{12} \, L_\odot$ \citep[][corrected for our
adopted cosmology]{priddey01}.
This quasar has also been detected in the molecular transition CO(5--4)
\citep{gui99} that provides an accurate determination of its redshift, which is
required for the search of the [CII] within the narrow band offered by
current submm receivers.
We have used the Atacama Path Finder Experiment (APEX)\footnote{APEX is
a collaboration between the Max-Planck-Institut fur Radioastronomie,
the European Southern Observatory, and the Onsala Space Observatory.}
to observe the [CII] line in BRI 0952-0115. In this letter we report the
successful detection of the line. We discuss some possible implications
of our detection
on the physics of high redshift galaxies and on the detectability of the
[CII] line in general in high redshift sources.
In this letter we assume the concordance $\Lambda$-cosmology with
$\rm H_0=71~km~s^{-1}Mpc^{-1}$, $\rm \Omega _{\Lambda}=0.73$ and
$\rm \Omega _{m}=0.27$ \citep[][]{spergel03}.

\section{Observations and results}

BRI 0952-0115 (RA(J2000)$=$09:55:00.1, DEC(J2000)$=$-01:30:07.1)
was observed with the Swedish Heterodyne Facility Instrument
\citep[SHFI,][]{vassilev08} on the APEX telescope
in six observing runs,
from November 1$^{st}$ to December 18$^{th}$ 2008 (a short run
on October 23$^{rd}$ was totally discarded because of bad weather conditions),
for a total of about
22 hours of observations, including sky observations and overheads,
resulting in 5 hours of net on-source integration.
The weather conditions were generally very good with
precipitable water vapor $\rm 0.25 < pwv < 1.0 ~mm$, the best quality
data coming from the run on December 17$^{th}$ when we had $\rm pwv < 0.5 ~mm$
for 6 hours. We used the APEX--2 receivers tuned to 349.77 GHz
(both polarizations), which is the expected frequency of the
[CII]157.741$\mu$m line at the
redshift z=4.4337 provided by the CO(5--4) detection \citep{gui99}.
At this frequency the 1~GHz band of the back-end translates
into a velocity coverage of about 860~km/s.
In the last run (December 18$^{th}$) we also observed with the frequency
of the two receivers offset by $\pm$200~km/s, to check the effects of possible
instrumental artifacts.
Observations were done in wobbler switching mode, with a symmetrical azimuthal
throw of 20$''$
and a frequency of 0.5~Hz. Pointing was checked on the nearby source
IRC+10216 every 1-2 hours, and found to be better than 3$''$
(with a beam size of 18$''$).
The focus was checked on Saturn,
especially after Sunrise when the telescope deformations are largest.

The data were analyzed by using CLASS (within the GILDAS-IRAM package).
One of the main problems in searching for relatively broad lines (a few
hundred km/s) with single dish telescopes are the spectral
baseline instabilities. As a consequence,
we visually inspected all individual scans and removed
those with clear baseline instabilities. We found that the analysis further
benefits of additional purging of scans that show a rms higher than a given
threshold. This technique removes additional
baseline instabilities which are not readily seen in the visual inspection
of individual scans. The lower the rms threshold, the lower the
``noise'' introduced by spurious baselines is, but an increasing fraction of
scans is discarded causing the Poisson noise to increase. We found that
an rms threshold of 130~mK optimizes the noise in the spectra (which
leaves about 1.6 hours of on-source integration), although
the results do not change significantly as long as the rms threshold
is below about 145~mK (which would include 60\% of the observing time).
To show that the final result does not depend
significantly on the scans selection (as long as the really bad ones
are removed) we have also separately combined all scans of the best
run, on December 17$^{th}$, without any rms rejection.
The individual scans were aligned in frequency and
averaged together.
A linear baseline was subtracted
to the final spectrum by interpolating the channels in the velocity
ranges $\rm -400<v<-200~km/s$ and $\rm 200<v<400~km/s$.

   \begin{figure}
   \centering
   \includegraphics[width=8truecm]{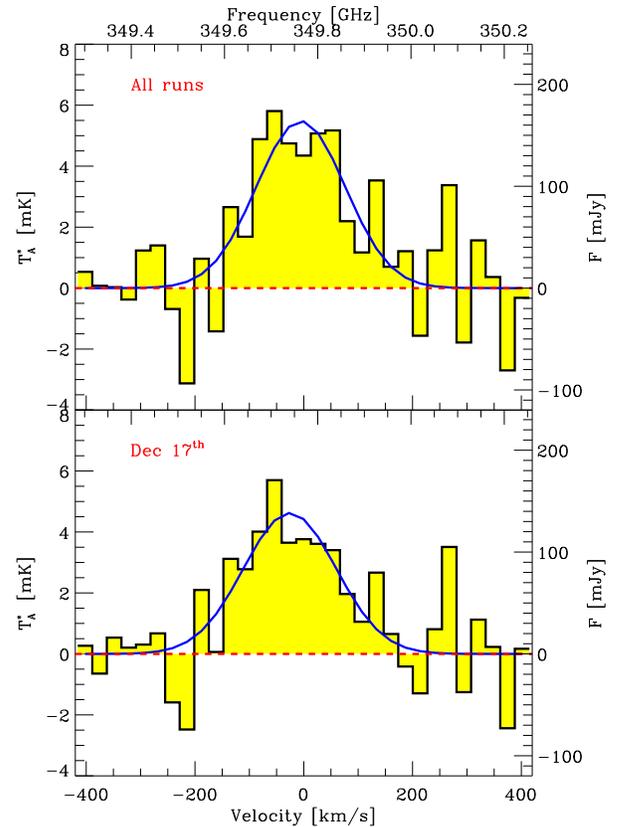}
      \caption{Spectrum of the [C{\sc II}] 157.74~$\rm \mu m$ emission line 
in the quasar BRI 0952-0115 at $z=4.4337$ shown with a velocity
resolution of $\rm 27 \, km \, s^{-1}$ (see text for details).
The blue curve shows the
gaussian fit to the line profiles (see Table~\ref{obs}).
              }
         \label{cii}
   \end{figure}

Fig.~\ref{cii} shows the resulting spectrum smoothed to a spectral resolution of
27~km/s. The top panel is for the combination of all runs with
the threshold rms$<$130~mK, while the bottom panel shows the combination
of the Dec.~17$^{th}$ run only (the best weather run),
without any rms rejection.
The [CII] line is clearly detected, with a significance of 7$\sigma$.
We also verified that if we split the observations in two halves the line
is detected independently in each of them with a significance higher than
about 5$\sigma$.

The [CII] line was fitted with a single gaussian. The resulting line
parameters are reported in Table~\ref{obs}, and compared
with the CO(5--4) line detected by \cite{gui99}. 
The [CII] line center and width are fully consistent with those of the CO line.

\begin{table*}
\caption{Properties of the [CII] line observed toward BRI 0952-0115 
compared with the CO(5--4) line observed by \cite{gui99}.}
\label{obs}     
\begin{tabular}{l c c c c c c} 
\hline\hline                
Line  & $\rm \nu _{rest}$ & $\rm \nu _{obs}$ & $z_{\rm line}$ & 
   $\rm \Delta V _{FWHM}$ & I & log(L)$^a$ \\
      & \multicolumn{2}{c}{[GHz]} &   &  [$\rm km~s^{-1}$] &
       [$\rm Jy~km~s^{-1}$] & [$\rm L_{\odot}$]\\
\hline                   
 [CII] ($\rm ^2P_{3/2}-^2P_{1/2}$) & 1900.54 & 349.776 & 4.4336$\pm$0.0003 &
       193$\pm$32 & 33.6$\pm$4.9 & 9.66$\pm$0.25 \\
 CO (5--4) & 576.2679 & 106.055  & 4.4337$\pm$0.0006 &  230$\pm$30  &
   0.91$\pm$0.11  & 7.58$\pm$0.25 \\
\hline
\end{tabular}\\
$^a$ Log of the
luminosity corrected for a lensing magnification of $\mu = 4.5$, while
the error reflects the possible range of magnifications $\mu =2.5-8$ (see
Appendix).
\end{table*}

\section{Discussion}

The previous two [CII]158$\mu$m detections at high redshift
\citep{maiolino05,iono06} were obtained in HyperLuminous Infrared Galaxies
($\rm L_{FIR}>10^{13}~L_{\odot}$). The observations
presented in this paper toward BRI 0952-0115 provide the first [CII]
detection at high redshift in a galaxy with luminosity
$\rm L_{FIR}<10^{13}~L_{\odot}$.
A striking result is the strong line flux, 5--8 times higher than
previous [CII] detections at high-z, with a peak intensity of 160~mJy.
Such a strong flux is partly due to the lensing amplification and
partly to the lower luminosity distance with respect to previous [CII]
detections.

The $\rm L_{[CII]}/L_{CO(5-4)}$ ratio is about five times higher
than inferred for the other two high-z sources with [CII] detection
\citep{maiolino05,iono06}. However, the latter two sources are actually
``CO--overluminous'', as shown by \cite{curran08}, and
anyhow the observed difference is well within
the spread between [CII] and CO luminosities observed
in local LIRGs \citep{curran08}.
A direct comparison with local galaxies is not easy, since most of local
galaxies have data for lower CO transitions, and the large variations
of the CO excitation, especially among high-z sources, makes it difficult
to translate the CO(5--4) luminosity into a CO(1--0) luminosity
\citep[e.g.][]{weiss07}.

   \begin{figure}
   \centering
   \includegraphics[width=9truecm]{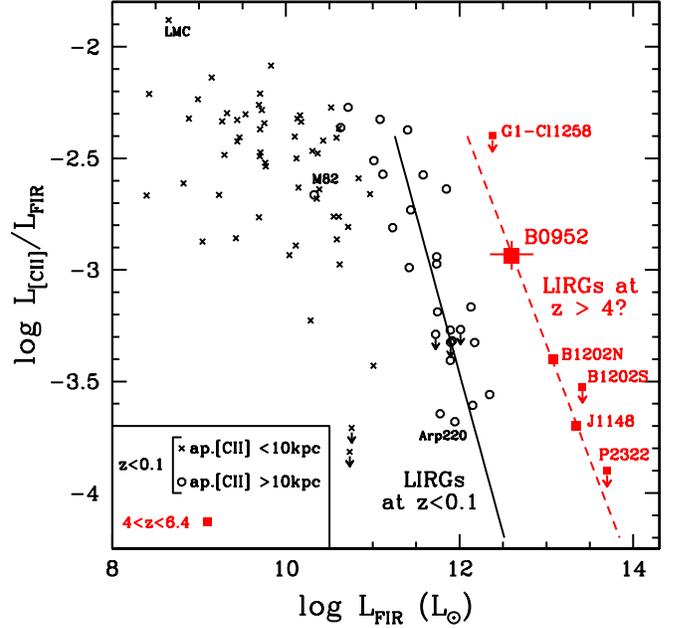}
      \caption{
$\rm L_{[CII]}/L_{FIR}$ ratio versus $\rm L_{FIR}$ for
normal and starburst local galaxies (black crosses and circles) and for high-$z$ sources
(red squares) \citep[references are given
in the text and in ][]{maiolino05}. The three high redshift [CII]
detections, including the one reported here (big symbol),
are labelled with the source name. The horizontal "errorbar" for
BRI 0952-0115 reflects the possible range of lensing magnification factors
(see Appendix).
Black crosses indicate galaxies for which the
aperture 
      to measure [CII] (ISO--LWS) is smaller than 10~kpc, and therefore
      it may sample a smaller region relative to the beam (IRAS) used for
      the far-IR flux.
The black solid line shows a linear fit (of the
logarithmic quantities) to the data of
local Luminous Infrared Galaxies (LIRGs, $\rm L_{FIR}>10^{11}~L{\odot}$).
The red dashed line is a linear fit to the detections at high redshift.
              }
         \label{cii_fir}
   \end{figure}

Probably the most important result of our observations is
that the [CII] line in BRI 0952-0115 is very strong relative
to its far-IR luminosity when compared to other luminous sources.
This is illustrated in Fig.~\ref{cii_fir},
which shows the $\rm L_{[CII]}/L_{FIR}$ ratio versus $\rm L_{FIR}$, both
for local galaxies (black crosses and open circles) and high-z galaxies
(red squares). Black crosses indicate galaxies for which the aperture
used to measure [CII] (generally ISO-LWS) is smaller than 10~kpc, and therefore
it may sample a smaller region relative to the beam (generally IRAS) used for the far-IR
flux (hence the $\rm L_{[CII]}/L_{FIR}$ ratio may not be reliable for
some of these sources). Local galaxies (black symbols) show the well known
drop of the $\rm L_{[CII]}/L_{FIR}$ ratio at luminosities
$\rm L_{FIR} > 10^{11}-10^{11.5}~L_{\odot}$.
In BRI 0952-0115 we measure a ratio $\rm log(L_{[CII]}/L_{FIR})=-2.93$
that is
higher than observed in local ULIRGs, which are mostly in the range
$\rm -3.7< log(L_{[CII]}/L_{FIR}) < -3.2$. More interesting is the
comparison of the trends of $\rm L_{[CII]}/L_{FIR}$ in local and
high-z galaxies. The black solid line in Fig~\ref{cii_fir} is a linear
fit (of the logarithmic quantities) to the local galaxies with 
$\rm L_{FIR}>10^{11} ~L_{\odot}$ ($\sim$ Luminous Infrared Galaxies, LIRGs,
see footnote \ref{foot_firdef}),
which describes the steep drop of $\rm L_{[CII]}/L_{FIR}$ at high luminosities
observed in local galaxies, in the form
\begin{equation}
\rm log \left( \frac{L_{[CII]}}{L_{FIR}}\right) _{local} =
-1.43~log(L_{FIR,12}) -3.47
\end{equation}
where $\rm L_{FIR,12}$ is the far-IR luminosity in units of
$\rm 10^{12}~L_{\odot}$.
The extrapolation of the black line to high luminosities heavily
underpredicts the $\rm L_{[CII]}/L_{FIR}$ ratio observed in high-z sources
(red squares). The high $\rm L_{[CII]}/L_{FIR}$ ratio observed in
BRI 0952-0115, along with the $\rm L_{[CII]}/L_{FIR}$ observed in the
other two high-z sources at higher luminosities, suggest that the [CII]
emission is enhanced in high-z galaxies relative to their local counterparts
of the same infrared luminosity.
The red dashed line in Fig.~\ref{cii_fir} is a linear fit
to the high-z [CII]-detected galaxies, in the form
\begin{equation}
\rm log \left( \frac{L_{[CII]}}{L_{FIR}}\right) _{high-z} =
-1.0^{+0.3}_{-0.6}~log(L_{FIR,12})
-2.3^{+0.7}_{-0.3}
\end{equation}
(where the errors on the coefficients are dominated by the uncertainty
on the magnification factor of BRI 0952-0115, see Appendix).
At a given infrared luminosity the
offset between the local (black-solid) and high-z (red-dashed) best fit lines
corresponds to an enhancement of the [CII] luminosity
by more than an order of magnitude in high-z galaxies.
Clearly the result depends on the actual magnification factor of
BRI 0952-0115, whose intrinsic $\rm L_{FIR}$ has a strong leverage on the
actual slope of the relation at high redshift (red dashed line). However,
even in the case of the highest possible
magnification factor ($\mu =8$, corresponding
to the leftmost end of the horizontal red bar in Fig.~\ref{cii_fir})
the evidence for [CII] enhancement, relative to local galaxies
of the same infrared luminosity, is still strong.

This result has still low statistical significance, since it is based
only on three high-z objects. Moreover, the observed trend
may result from selection effects, in the sense that high-z [CII]
non-detections may have not been published, hence the three published detections
may trace the upper envelope of a wider distribution. Yet, it is worth noting
that the three high-z [CII] upper limits available so far
\citep[][P. Cox priv. comm.]{marsden05,iono06}\footnote{BR 1202-0725
was included in
\cite{maiolino05} as an upper limit (from Benford et al., in prep.),
but subsequently \cite{iono06} detected [CII] in the northern component of
the source, while the southern component remained undetected. Both
components are consistent with the high-z relation shown in Fig.~\ref{cii_fir}.}
are consistent
with the high-z [CII]-enhanced scenario, as illustrated
in Fig.~\ref{cii_fir}.
If the observed effect is confirmed with a larger sample of high-z galaxies
the implied consequences would be quite important, both for
the physics of high-z galaxies and for the use of [CII]
to search and investigate high-z galaxies in future surveys.

In the following we speculate on the possible origin of the enhanced [CII]
emission in high-z galaxies {\rm relative to local galaxies of the
same luminosity}. The offset observed in
Fig.~\ref{cii_fir} between local and high-z galaxies
cannot be ascribed to an additional contribution to $\rm L_{FIR}$
due, for instance, to the AGN hosted in these high-z systems (and anyhow
present also in several of the local galaxies). Indeed, a varying
$\rm L_{FIR}$ contribution by AGNs would move objects nearly parallel to the
black solid line. Moreover, the far-IR luminosity of (sub-)mm bright quasars
(such as the ones discussed here) is generally found to
be mostly powered by star formation \citep{lutz07,lutz08}.

Another possibility is that these high-z galaxies are 
characterized by a lower metallicity of the ISM. Observationally, local
low metallicity galaxies tend to show enhanced [CII]158$\mu$m emission
\citep{rubin09,pog95,israel96,madden00}. The effect
is apparent from the location of LMC in Fig.~\ref{cii_fir} \citep{rubin09}.
Probably the enhanced
[CII] emission in low metallicity galaxies is due to
the lower dust content (hence
lower dust attenuation to UV photons), which makes the C$^{+}$ emitting region
larger, and also makes the far-IR emission lower
\citep{rubin09}. Regardless of 
the physical origin of the effect, if high-z galaxies are characterized
by a reduced metallicity, this may enhance their [CII] emission similarly to
local low metallicity galaxies.
High-z star forming galaxies are indeed
observed to have lower gas metallicities than
local galaxies \citep{maiolino08b}. For galaxies hosting quasars, as
the ones investigated here, the
situation is more complex. Various studies have found that the
metallicity in the Broad Line Region (BLR) of high-z quasars is very high
(several times solar) and does not evolve with redshift
\citep{juarez09,jiang07,nagao06a}. However, the BLR is a very tiny region
($<$1pc) in quasar nuclei, which is probably not representative
of the ISM in the host galaxy, and may undergo quick enrichment with just a few
supernova explosions \citep[see detailed discussion in][]{juarez09}.
A few studies have investigated the metallicity on the larger scales
($\sim$100pc--10kpc) of the Narrow Line Region (NLR) in high-z AGNs
\citep{nagao06b,matsuoka09,vernet01,debreuck00,humphrey08}. These studies are currently limited to z$<$4
(i.e. not yet overlapping with our high-z [CII]-sample). However the
inferred NLR metallicities are much lower than in the BLR, about solar or
sub-solar,
which is indeed less than the metallicity observed in local massive galaxies
\citep{maiolino08b}.

Regardless of the physical origin of the [CII] enhancement in
high-z galaxies, if confirmed this effect would have important implications
for the planning of future surveys at high-z as well as for the
development of submm/mm facilities. Our result would imply
that, at least at high infrared luminosities, the [CII] line in high-z
galaxies is about an order
of magnitude stronger than previously expected based on local templates.
The [CII] line is probably a much more powerful cosmological tool to detect
and characterize high-z galaxies than previously thought.


\begin{acknowledgements}
We thank the anonymous referee for very helpful comments
and M. Bartelmann for useful discussions.
     We are grateful to the ESO-APEX staff for their support and help
	 during the execution of the project. 
	  \end{acknowledgements}

\appendix
\section{The lensing magnification factor}

Although $\rm L_{[CII]}/L_{CO}$ and $\rm L_{[CII]}/L_{FIR}$ most
likely do not depend on the magnification factor, since [CII], CO and
FIR are emitted from the same regions of the galactic gaseous disk, the
inferred intrinsic luminosities, and in particular $\rm L_{FIR}$ do
depend on the magnification factor. Therefore, especially for what concerns
the $\rm L_{[CII]}/L_{FIR}$ versus $\rm L_{FIR}$ diagram, it is
important to address the issue of the magnification lensing factor in detail.

\cite{gui99} provided
only a rough estimate of the magnification factor ($\rm \mu\sim 4$).
\cite{lehar00} performed a more detailed lensing analysis by
using different mass models for the lensing galaxy along with HST images (although
they do not directly provide the associated magnification factors).
The positions and the flux ratio
of the two quasar images are well reproduced by a Singular-Isothermal-Ellipsoid model
(SIE). A constant M/L model, where the lensing mass distribution matches the light
distribution of the observed galaxy, plus an external shear $\gamma$ (due to galaxies
surrounding the lens) also provides a good fit to
the data. Details on the best fit parameters describing the lensing models can be found
in Tab.~5 of \cite{lehar00}. Based on the image constraints, it is impossible
to differentiate between the SIE and the M/L models. We used the public software {\it
GRAVLENS} and its application {\it LENSMODEL} \citep{keeton01} to repeat the fits by
\cite{lehar00} and to reconstruct their lens models. These were used to estimate the
magnification factors at the positions of the two quasar images. The lens is a typical
early-type galaxy, whose light distribution can be described by means of a De Vaucouleurs
profile.  The two models predict significantly different magnification factors. Assuming
a point source, the SIE and the $M/L+\gamma$ models give total magnification factors
$\mu=8$ and $\mu=2.5$, respectively. The large difference is caused by the substantially
different slope of the two density profiles, being the SIE much steeper than the De
Vaucouleurs profile.  Without an external shear, the SIE model may be
overestimating the lens convergence and therefore
the magnification. However, since no
correlation is expected between the external shear and the environment of the lens
galaxy, it is hard to estimate the possible bias \citep[see the discussion in ][]{lehar00}.
In the $M/L+\gamma$ model, even a modest external shear strongly perturbs the
lens. In general, steeper profiles are favored by observations of strong lensing systems
\citep[e.g.][]{rusin03,treu04}. Therefore we can assume that the
magnification factor predicted by the $M/L+\gamma$ model is a lower limit to the true
value.

By using the result of the fit of the quasar double image, we attempted to estimate by how
much the magnification factor changes for an extended source. This is
important in our case since the [CII] and the FIR emissions come from an extended
region of the gaseous disc of the quasar host galaxy. For instance, in the case
of J1148+5251 at z=6.41, high angular resolution images show that [CII] is distributed
in a region of about 1.5~kpc in diameter around the quasar \citep{walter09}.
By replacing, on the source plane, the point-like source
(the quasar nucleus observed in the optical images)
with a circular disk (the [CII]-FIR emitting region)
of radius $0.2"$ (equivalent to $\sim 1\;h^{-1}$kpc at $z=4.43$) centered at the location
of the un-lensed quasar (as it results from the
modeling of its point images), and mapping it on the lens plane, we find that the total
magnification factors change by $\lesssim 0.5$ for both the SIE and the $M/L+\gamma$
models. The magnification factor does not change significantly when varying the size of
the [CII]--FIR disk as long as it is less than 2~kpc in radius.
However, we note that the morphology of the extended images is very different for
the two models. In particular, the SIE model produces an Einstein ring-like image, while the
$M/L+\gamma$ model produces an extended asymmetric arc and a counter image. Therefore,
high angular resolution observations of the [CII] and FIR emission would allow us to
discriminate which of the two lens models is more appropriate.

Summarizing, the magnification factor ranges between $\mu =2.5$ and $\mu=8$,
mostly depending on the lens model.
In the paper we assume, as a reference, a magnification
factor $\mu = 4.5$ (the mean value of the two extreme magnification
factors in log), although we discuss the implications of the
wide range of possible magnification factors.
For what concerns the intrinsic far infrared luminosity, the observed
value $\rm L_{FIR}=1.7~10^{13}~L_{\sun}$ inferred by \cite{priddey01} (and
corrected for our assumed cosmology)\footnote{Note that moderate uncertainties
in the observed $\rm L_{FIR}$ move the source essentially along the
red dashed line in Fig.~\ref{cii_fir}.}
, translates into an intrinsic,
magnification-corrected far-infrared luminosity of
$\rm L_{FIR} = 4^{+3}_{-2}~10^{12}~L_{\sun}$, where the errors reflect the
range of possible magnification factors.

\end{document}